\documentclass[french,american]{article}
\usepackage[T1]{fontenc}
\usepackage[utf8]{inputenc}
\usepackage{babel}
\usepackage{amsmath}
\usepackage{amsthm}
\usepackage{amssymb}
\usepackage[pdfusetitle,
bookmarks=true,bookmarksnumbered=false,bookmarksopen=false,
breaklinks=false,pdfborder={0 0 1},backref=false,colorlinks=false]
{hyperref}

\makeatletter

\theoremstyle{plain}
\newtheorem{theorem}{Theorem}[section]

\newtheorem{proposition}[theorem]{Proposition}
\newtheorem{corollary}[theorem]{Corollary}

\theoremstyle{definition}
\newtheorem{definition}[theorem]{Definition}

\newtheorem{remark}[theorem]{Remark}

\theoremstyle{remark}

\theoremstyle{plain}
\newtheorem*{definition*}{Definition}
\newtheorem*{theorem*}{Theorem}
\newtheorem*{proposition*}{Proposition}
\newtheorem{fact}[theorem]{Fact}
\newtheorem{question}{Question}

\author{
	Djamel Eddine AMIR \\
	\normalsize\textit{Université Paris-Saclay, CNRS, LISN, 91400, Orsay, France} \\
	\normalsize\href{mailto:amir@lisn.fr}{amir@lisn.fr} 
	\normalsize\href{https://amirdjameleddine.github.io/}{https://amirdjameleddine.github.io/}	
	\and
	Benjamin HELLOUIN de MENIBUS \\
	\normalsize\textit{Université Paris-Saclay, CNRS, LISN, 91400, Orsay, France} \\
	\normalsize\href{mailto:hellouin@lisn.fr}{hellouin@lisn.fr} 
	\normalsize\href{https://www.lisn.fr/~hellouin/}{https://www.lisn.fr/~hellouin/}
}

\date{}

\newcommand{\Z}{\mathbb{Z}}
\newcommand{\N}{\mathbb{N}}
\newcommand{\A}{\mathcal A}
\renewcommand{\L}{\mathcal L}
\DeclareMathOperator{\Orb}{Orb}
\DeclareMathOperator{\Per}{Per}

\title{Minimality and computability of languages of G-shifts}

\begin{document}
	
	\maketitle

\begin{abstract}

Motivated by the notion of strong computable type for sets in computable
analysis, we define the notion of strong computable type for $G$-shifts,
where $G$ is a finitely generated group with decidable word problem. A
$G$-shift has strong computable type if one can compute its language from
the complement of its language.
We obtain a characterization of $G$-shifts
with strong computable type in terms of a notion of minimality with
respect to properties with a bounded computational complexity.
We provide a self-contained direct proof, and also explain how this characterization
can be obtained from an existing similar characterization for sets by Amir and Hoyrup, and discuss its connexions with results by Jeandel on closure spaces.
We apply this characterization to
several classes of shifts that are minimal with respect to specific
properties.
This provides a unifying approach that not only generalizes many
existing results but also has the potential to yield new findings
effortlessly. In contrast to the case of sets, we prove that strong
computable type for G-shifts is preserved under products. We conclude by
discussing some generalizations and future directions.
\end{abstract}

\section{Introduction}

Shifts, or shift spaces, are sets of colourings of an infinite regular grid (also called configurations) submitted to local constraints usually given as forbidden patterns. In 1961, Wang first studied colourings of the two-dimensional square grid with finitely many constraints (shifts of finite type or \emph{SFT}) to study some fragments of predicate calculus \cite{wang1961proving}. Undecidability phenomena were proved soon after, the first one being the undecidability of the so-called seeded Domino problem \cite{kahr1962entscheidungsproblem}: given a partial colouring, can it be extended to a full configuration that satisfies the constraints? The set of such partial colourings is called the \emph{language} of the shift, so this result means that there are SFT with uncomputable language. This result was later extended to other grids, in particular Cayley graphs of finitely generated groups \cite{bitar2023contributions}, which is the general setting for this article.

Consequently, understanding which properties
make the language of a shift computable or not remains a fundamental question.
A well-known folklore result states that the language of a minimal
SFT, that is, a shift that contains no other non-empty subshift, is computable. However, later research showed that the finite-type
assumption is not necessary: it suffices to be able to enumerate all
patterns that do not appear in the shift (a notion known as effectiveness)
for the result to hold (see \cite{jeandel2019characterization,jeandel2017enumeration,delvenneblondel}).
Multiple other results point to the same phenomenon, where being
able to enumerate the complement of the language is enough to compute
the language itself, under various assumptions that can be seen as a form of minimality: the subshift satisfies a property $P$ that none of its subshifts satisfies.

\paragraph*{Similar results in other areas}

The connection between minimality and decidability also extends to
problems in group theory, combinatorics, and computable analysis (see
for example \cite{higman1961subgroups,lyndon1977combinatorial,Miller02}).
Jeandel attempted to develop a unified theory for groups, shifts and
combinatorics \cite{jeandel2017enumeration}; we discuss the relationship with our results in Section~\ref{sec:quasivar}.

In the setting of computable topology and descriptive complexity, a similar
phenomenon occurs where some minimality assumption ensures that a description of a set "from the outside"
is enough to provide a description "from the inside". Notions of strong computable type and minimality for sets were defined
in \cite{AH22c} as follows.
\begin{definition*}
A \textbf{set }has \textbf{strong computable type} if one can compute
the set from its semi-computable information (a description of its complement).

A \textbf{set} is \textbf{minimal} for some property if it satisfies
this property but no proper subset of it satisfies the property. 
\end{definition*}
A characterization of sets which have strong computable type related
to minimality was given in \cite{AH22c}.
\begin{theorem*}[\cite{AH22c}]
A \textbf{set} has \textbf{strong computable type} if and only if
it is minimal satisfying some property with bounded computational complexity ($\Sigma_2^0$-computable).
\end{theorem*}

\paragraph*{Our approach}

In this article, motivated by the above results for sets, we define
analogous notions of strong computable type and minimality for $G$-shifts,
where $G$ is a finitely generated group with decidable word problem.
To simplify, $G$-shifts will be called shifts.
\begin{definition*}
A \textbf{shift }has \textbf{strong computable type} if one can compute
its language from the complement of its language.

A \textbf{shift} is \textbf{minimal} for some property if it satisfies
this property but no proper subshift of it satisfies the property.
\end{definition*}
We characterize shifts with strong computable type using minimality,
and show that this is implied by the characterization of strong computable type for sets. This motivates the question of classifying shifts according to their computability.

In addition to creating new connections between symbolic dynamics
and computable analysis, positive results in this direction will motivate
further efforts to find more general theorems that unify the two fields.

\paragraph*{Our results}

We prove the following main theorem.
\begin{theorem*}[Theorem \ref{thm:main}]
A \textbf{shift} has \textbf{strong computable type} if and only
if it is minimal for some property with bounded computational complexity ($\Sigma_2^0$-computable).
\end{theorem*}
We apply our theorem to several classes of shifts which are minimal
for some specific properties. This provides a unifying approach, as
it not only implies many existing results but also introduces new
findings effortlessly. 
\begin{proposition*}[Section \ref{subsec:Application}]
The following classes of shifts have strong computable type.
\begin{itemize}
\item Minimal shifts,
\item Entropy-minimal shifts with a left-computable entropy,
\item $P$-isolated shifts,
\item Infinite-minimal shifts (periodic-minimal shifts and quasi-minimal
shifts).
\end{itemize}
\end{proposition*}
More details and precise definitions will be provided in the following
sections, along with an appendix for clarifications. 

Many other natural questions arise, such as: can we obtain other analogous
results for shifts to those on computable type for sets (\cite{Phdamir2024,AH22,AH22c,AH23,AMIR2024109020,AMIR2025103611,CCAinvitedAMIR,čačić2024computableapproximationssemicomputablegraphs,Čelar2021})?
Furthermore, can we reverse the process by obtaining results for computable
type analogous to those on shifts?

\paragraph*{Sections are organized as follows:}

In Section \ref{sec:Preliminaries}, we give a minimal background
to understand the main results. In Section \ref{sec:Strong-computable-type},
we define the notion of strong computable type for shifts, state and
prove our main theorem, explain its relation with strong computable
type for sets and closure spaces, apply it to classes of shifts and study some properties of strong computable type for shifts. In Section \ref{sec:Generalization,-conclusion-and},
we discuss generalizations of our results and future directions. In
Appendix \ref{Appendix:Topology-on}, we provide more details about
topology and descriptive complexity. In Appendix \ref{Appendix:Another-proof-of},
we prove the relation between strong computable type for shifts and
sets.

\section{Preliminaries}\label{sec:Preliminaries}

To present our results, which are interdisciplinary in nature, as
they relate computability and symbolic dynamics to topology and descriptive
complexity, we first provide some basic concepts. However, for a more
comprehensive understanding, the reader should refer to the relevant
references provided throughout the text.

\subsection{Basics on groups}

We provide some classical basic concepts about groups,
particularly the notions of finitely generated groups and the word
problem. These results and their applications to shifts can be found in \cite{aubrun2018domino}.
\begin{definition}
An \textbf{alphabet} is a set. A \textbf{word }of \textbf{length}
$n\in\mathbb{N}^{*}$ on an alphabet $A$ is an element in the finite
product $\Pi_{1\leq i\leq n}A$, the \textbf{empty word} is of length
$0$. 

We denote by $A^{*}$ the set of all finite words on $A$.
\end{definition}

Now, we have the necessary ingredients to define key concepts for
groups.
\begin{definition}
Let $G$ be a group and $S\subseteq G$. For $w\in S^{*}$,
we denote the corresponding group element by $w_{G}$ (the evaluation
of $w$ as an element of $G$, where the product is the group
operation of $G$). 

$S$ is a \textbf{generating set} for $G$ if every element in $G$
can be written as a word in $S^{*}$. $G$ is \textbf{finitely generated}
if $S$ is finite.
\end{definition}

To simplify, we assume that generating sets are symmetric, that is, $g\in S\Rightarrow g^{-1}\in S$.

\begin{definition}
Let $G$ be a group that is finitely generated by $S$ with identity element $1_G$.

The \textbf{word problem} is the set $\{w\in S^{*}:w_{G}=1_{G}\}$.

$G$ has \textbf{decidable word problem} if this set
is decidable, i.e. there is an algorithm which given a word $w$,
decides on finite time whether $w_{G}$ equals $1_{G}$. This does not depend of
the chosen generating set.
\end{definition}

\subsection{G-shifts}

As outlined in the introduction, shifts of dimension $2$ can be defined
as sets of colorings of planes. Specifically, given an alphabet $A$
(a set of colors), a configuration corresponds to an element of $A^{\mathbb{Z}^{2}}$,
i.e., an assignation of colors $(a_{(i,j)})_{(i,j)\in\mathbb{Z}^{2}}$ to every coordinate in the plane, where each $a_{(i,j)}$ is
a color in $A$. A shift of dimension $2$ is a shift-invariant closed
set of configurations in $A^{\mathbb{Z}^{2}}$.

Similarly, this concept can be extended by indexing the colors with
elements of a group $G$ rather than $\mathbb{Z}^{2}$, i.e. configurations
$(a_{g})_{g\in G} \in A^G$, which allows us to define $G$-shifts (see \cite{ceccherini2010cellular}, \cite{aubrun2018domino}).
A configuration $(a_{g})_{g\in G}$ can be seen as a function
$G\rightarrow A$ sending $g$ to $a_{g}$. 

Now, let us give precise definitions for $G$-shifts.

Let $G$ be a group and let $A$ be a finite alphabet. The set $A^{G}=\{x:G\rightarrow A\}$ is called the \textbf{full shift},
and its elements are called \textbf{configurations}. $A^{G}$ can be endowed with the \textbf{pro-discrete} topology, which is
metrizable if $G$ is countable and computably metrizable if $G$
is finitely generated. For the latter case, see Appendix \ref{subsec:Topology-on-1} for more details.

\begin{definition}[$G$-shifts]

A \textbf{pattern} is an element $p\in A^{F}$, where $F\subseteq G$
is finite, and it determines the \textbf{cylinder} $[p]=\{x\in A^{G}:x|_{F}=p\}$.
Let $x:G\rightarrow A$ be a configuration, $p$ \textbf{appears}
on $x$ if there exists some $g\in G$ such that $gx\in[p]$, where
$gx:G\rightarrow A$ is the function sending $h\in G$ to $x(g^{-1}h)$.

A \textbf{$G$-shift} is a subset $X\subseteq A^{G}$ which is topologically
closed and \textbf{shift-invariant} i.e. for every $h\in G$ and every
$x\in X$, one has $hx\in X$. 
We denote by $\mathcal{S}(A^{G})$ the set of all non-empty shifts in $A^{G}$.

The \textbf{language} $\mathcal{L}(X)$ of a $G$-shift $X$ in $A^{G}$
(or more generally a set $X\subset A^G$) is the set of patterns $p: F\to A$ that \textbf{appear} in $X$, that is, such that $x|_F=p$ for some $x\in X$.
Its complement is denoted $\mathcal{L}^{c}(X)$. 

A $G$-shift $X$ is \textbf{of finite type (SFT)} if it can be defined
by a finite set $F$ of forbidden patterns, in the sense that a configuration
$x$ is in $X$ if and only if no pattern from $F$ appears in $x$.
\end{definition}

To simplify, we call $G$-shifts just shifts, the underlying group
will be clearly known. In the literature, shifts may be called subshifts
since every shift is a subshift of the full shift.

If $X$ is a shift, then a pattern $p$ appears in some $x\in X$
if and only if $X\cap[p]\neq\emptyset$ (this will be used in the
proofs).

Note that the set of cylinders $[p]$ is a clopen sub-basis of the pro-discrete
topology on $A^{G}$. 

\subsection{Effectiveness of G-shifts}

Let $G$ be a finitely generated group with decidable word problem,
$S\subseteq G$ be a finite generating set and $A$ be a finite alphabet.
To define notions of effectiveness on $G$-shifts (see \cite{Hochman2009}),
let us define some classical notions in computability theory.
\begin{definition}
A subset $I\subseteq\mathbb{N}$ is \textbf{computably enumerable
(c.e.)} if there exists an algorithm that runs forever and enumerates
all elements of $I$; equivalently, there is an algorithm that \textbf{semi-decides} $I$, that is, given $i\in \mathbb{N}$, the algorithm will enumerate $i$ after a finite time if $i\in I$ and keeps running forever otherwise.

A set is \textbf{co-computably enumerable (co-c.e.)} if its complement
is computably enumerable. 

A set is \textbf{computable (decidable)} if it is both c.e. and co-c.e..
\end{definition}

Let $p\in A^{F}$ be a pattern, that is, a function $F\rightarrow A$, for some finite set $F\subseteq G$. Let $f:S^{*}\rightarrow G$
be the function sending a word $w$ to the corresponding group element
$w_{G}$.
When manipulated by an algorithm, the pattern $p$ is represented by a function $p':W\rightarrow A$, with $W\subset S^*$ finite, such that $p\circ f|_W = p'$ and $W\cap f^{-1}(i)\neq \emptyset$ for all $i\in F$.
The set $\mathcal{C}$ of all such functions is countable as they
are functions from a finite set of words to a finite alphabet:
\[
\mathcal{C}=\{p\circ f_{F}:p\in A^{F},F\subseteq G\text{ is finite}\}.
\]
Note that $\mathcal{C}$ can be effectively enumerated (using an algorithm),
given $S$ and the fact that $G$ has  decidable word problem.

Hence the notions of c.e., co-c.e. and computable sets can be extended
from subsets of $\mathbb{N}$ to subsets of $\mathcal{C}$.

\begin{definition}[Effectively closed shifts]
A shift $X\subseteq A^{G}$ is \textbf{effectively closed} if it has a co-computably enumerable language.
\end{definition}

Such shifts are sometimes simply called \emph{effective}.

\begin{fact}[Proposition~2.1 in \cite{aubrun2018domino}]\label{fact:effective-altdef}
Equivalently, a shift is effectively closed if it can be defined by a computably enumerable set of forbidden patterns.
\end{fact}
In particular, an SFT is effectively closed because a finite set of forbidden patterns is computably enumerable.

This terminology comes from the fact that a shift is effectively closed if
and only if it is topologically effectively closed (see \cite{AUBRUN201735}
and Appendix \ref{subsec:Descriptive-complexity-for}). Since a shift
is always topologically closed, an effectively closed shift corresponds to
an effectively topologically closed shift.

\subsection{Descriptive complexity}

In this section, we define the notion of properties of shifts, their
descriptive complexity and minimal elements satisfying them.

Let $G$ be a finitely generated group with decidable word problem
and let $A$ be a finite alphabet. The set $\mathcal{S}(A^{G})$ of
all non-empty shifts in $A^{G}$ can be endowed with a computable
metric structure, see Appendix \ref{subsec:The-hyperspace-of}. 

When we say that an algorithm is given an enumeration as input, this formally means that the enumeration is given as an oracle to the algorithm (that is, the algorithm has access to it as a special input of infinite length).

A \textbf{property} is a subset of $\mathcal{S}(A^{G})$.
\begin{definition}[See Section~2 in \cite{cenzer99}]
\label{def:The-descriptive-complexity}The \textbf{descriptive complexity}
of a property $P$ can be defined as follows: a property is $\Pi_{1}^{0}$  if it is effectively closed. Equivalently (see Fact~\ref{fact:effclo<>p01}), there exists an algorithm such that for every $X\in\mathcal{S}(A^{G})$, given two enumerations of its language $\mathcal{L}(X)$ and of its complement $\mathcal{L}^{c}(X)$, the algorithm halts if and only if $X\notin P$ (it semi-decides whether $X\notin P$).

A property is $\Sigma_{1}^{0}$ if its complement is $\Sigma_{1}^{0}$
(the algorithm semi-decides whether $X\in P$). It is $\Sigma_{2}^{0}$
if it is a uniform union of $\Pi_{1}^{0}$ properties, that is, $P = \bigcup_{i\in\N}P_i$ and there is an algorithm that, given $i$, semi-decides if a shift $X$ is not in $P_i$.
\end{definition}

\begin{definition}
Let~$P\subseteq\mathcal{S}(A^{G})$ be a property and~$X\in\mathcal{S}(A^{G})$
be a non-empty shift. We say that~$X$ is~$P$\textbf{-minimal}
if $X\in P$ and, for every non-empty subshift $Y\subsetneq X$, $Y\notin P$.
\end{definition}

A subshift $Y$ of a shift $X$ can always be obtained by forbidding an additional set of patterns; that is, $Y = X\setminus\bigcup_{p\in P}\bigcup_{g\in G}g[p]$ for some set $P \subseteq \mathcal L(X)$, where for $p:F\to A$ we define $g[p]=\{gx\in A^{G}:x\in A^{G}\text{ and }x|_{F}=p\}$.

\section{Strong computable type for G-shifts}\label{sec:Strong-computable-type}

\subsection{Definition}

We introduce the notion of strong computable type for $G$-shifts,
which originates from a similar concept in computable analysis (see
Section \ref{subsec:SCTforsets}).

As mentioned in the introduction, an effectively closed shift has computable language if it satisfies additional properties, such as being minimal. Identifying such sufficient conditions is of particular interest.
We generalize this question as follows: given information on the co-language of a shift (semi-computable or "negative" information), when can we use it to compute its language? This corresponds to asking under which conditions there is an algorithm that enumerates $\mathcal L(X)$ when provided with any oracle $O$ that makes $\mathcal L(X)^c$ computably enumerable.

This definition is formalised as follows, where an \textbf{oracle} is viewed as a subset of $\mathbb{N}$.
\begin{definition}[Strong computable type for $G$-shifts]
Let $G$ be a finitely generated group with  decidable word problem,
$A$ be a finite alphabet and $X\subseteq A^{G}$ be a shift. 

$X$ has \textbf{strong computable type} if for every oracle $O\subset\mathbb{N}$,
the fact that $X$ is effectively closed relative to $O$ implies that the
language of $X$ is computable relative to $O$.
\end{definition}

This is equivalent, by Selman's theorem \cite{selman1971arith}, to the fact that $\mathcal L(X)$ is enumeration-reducible to $\mathcal L^c(X)$.

Note that for effectively closed shifts (in particular SFTs) having computable
language is equivalent to having strong computable type. 

There exist shifts that have strong computable type but do not have
computable language. Indeed, a shift $X$ which has a c.e. language
which is not co-c.e. clearly has strong computable type; consider for example the shift on $\{0,1\}^\Z$ obtained by forbidding all patterns $01^k0$ such that the $k$-th Turing machine doesn't halt.

This notion is interesting because it provides a characterization
that unifies various arguments, offers new results, and connects the
computability of languages to the descriptive complexity of topological
properties (see Section \ref{subsec:Subshifts-and-minimality}).

\subsection{Shifts and minimality}\label{subsec:Subshifts-and-minimality}

In this section, we prove our main theorem, which characterizes shifts
with strong computable type by the fact that they must satisfy a property
that makes them minimal. This is motivated by Theorem
4.7. in \cite{AH22c}.

\begin{theorem}[Main theorem]
\label{thm:main}Let $G$ be a finitely generated group with
 decidable word problem and let $A$ be a finite alphabet. Let $X\subseteq A^{G}$
be a non-empty shift. The following are equivalent.
\begin{enumerate}
\item $X$ has strong computable type,
\item There exists a $\Sigma_{2}^{0}$ property $P$ in $\mathcal{S}(A^{G})$
such that $X$ is $P$-minimal.
\end{enumerate}
\end{theorem}

\begin{remark}
From the proof of the implication $2.\Rightarrow1.$ below, we can see that Theorem~\ref{thm:main} holds when the property $P$ is only $\Sigma_2^0$ relative to $\mathcal L^c(X)$ (that is, the algorithm used to prove that $P$ is $\Sigma_2^0$ receives $\mathcal L^c(X)$ as an additional oracle, where $X$ is the fixed subshift in the theorem statement).
\end{remark}

\begin{proof}
By Selman's theorem, Condition $1.$ is equivalent to the fact that an effective procedure produces an enumeration of
$\mathcal{L}(X)$ from any enumeration of its complement $\mathcal{L}^{c}(X)$.

$1.\Rightarrow2$. Assume that there is an effective procedure (a
machine $M$) producing an enumeration of $\mathcal{L}(X)$ from any
enumeration of $\mathcal{L}^{c}(X)$, let us define some $\Pi_{1}^{0}$
(and hence $\Sigma_{2}^{0}$) property $P$, for which $X$ is $P$-minimal. 

Let $\varGamma$ be the set of all pairs $(\sigma,p)$ where $\sigma=\sigma_{0}\ldots\sigma_{k}$
is a finite sequence of patterns such that the machine $M$ outputs
the pattern $p$ after reading $\sigma$; clearly $\varGamma$ is
c.e.. 

Let $\mathcal{U}=\bigcup_{(\sigma,p)\in\varGamma}\mathcal{U}_{(\sigma,p)}$
with
\[
\mathcal{U}_{(\sigma,p)}=\{X'\in\mathcal{S}(A^{G}):\forall i\leq k,X'\cap[\sigma_{i}]=\emptyset\text{ and }X^{\prime}\cap[p]=\emptyset\}.
\]

Note that a c.e. union of $\Sigma_{1}^{0}$ properties is a $\Sigma_{1}^{0}$
property. $\mathcal{U}$ is a $\Sigma_{1}^{0}$ property because $\varGamma$
is c.e. and for every $(\sigma,p)\in\Gamma$, $\mathcal{U}_{(\sigma,p)}$
is a $\Sigma_{1}^{0}$ property (given an enumeration of $\mathcal{L}^{c}(X')$
one can semi-decide whether the $\sigma_{i}$'s and $p$ are in $\mathcal{L}^{c}(X')$,
equivalently whether the corresponding cylinders do not intersect
$X'$). 

Let $P$ be the complement of $\mathcal{U}$, it is hence a $\Pi_{1}^{0}$
property. Let us prove that $X$ is $P$-minimal.

$X$ is in $P$ because otherwise the machine $M$ fails on $X$,
i.e. after reading a finite sequence of patterns $\sigma$ that are
in $\mathcal{L}^{c}(X)$ the machine outputs a pattern $p$ that is
in $\mathcal{L}^{c}(X)$ (which contradicts our assumption). 

Now, let $Y\subsetneq X$ be a shift and $p \in \mathcal L(X)\setminus \mathcal L(Y)$. We have $Y\notin P$ because the machine eventually outputs~$p$ after reading some finite sequence~$\sigma$, which implies that $Y\in\mathcal{U}_{(\sigma,p)}$.

$2.\Rightarrow1$. Assume that~$P$ is $\Sigma_{2}^{0}$ and that~$X$
is~$P$-minimal. Hence, $P=\bigcup_{i\in I}P_{i}$ with $P_{i}\in\Pi_{1}^{0}$ for all $i$. It is easy to see
that there exists some $i_{0}$ such that $X$ is $P_{i_{0}}$-minimal. 

By Fact~\ref{fact:PinV<>PinupV}, we can suppose that $P_{i_{0}}$ is $\Pi_{1}^{0}$
in a stronger sense: there exists an algorithm such that for every
$Y\in\mathcal{S}(A^{G})$, given an enumeration of the
complement of its language $\mathcal{L}^{c}(Y)$, the algorithm halts
if and only if $Y\notin P$; see Appendix \ref{subsec:Another-definition-of} for more details.

Given an enumeration of $\{q:[q]\cap X=\emptyset\}$ (that is,
an enumeration of $\mathcal{L}^{c}(X)$), we need to enumerate $\mathcal L(X)$. Note that~$p\in \mathcal L(X)$ iff $X' = X\setminus\bigcup_{g\in G}g[p]$
is a proper subshift in~$X$; as~$X$ is~$P_{i_{0}}$-minimal,
it is equivalent to $X'\notin P_{i_{0}}$.

$X'$ can be defined by the set of forbidden patterns $\mathcal L^c(X)\cup\{p\}$. There is an algorithm that, for any shift $Y$, computes an enumeration of $\mathcal L^c(Y)$ from any enumeration of a set of forbidden patterns that defines $Y$; this can be proved exactly as Fact~\ref{fact:effective-altdef} relative to an oracle. In particular, from an enumeration of $\mathcal L^c(X) \cup \{p\}$ we can compute, uniformly in $p$, an enumeration of $\mathcal L^c(X')$.

Since $P_{i_{0}}$ is~$\Pi_{1}^{0}$, from an enumeration of $\mathcal{L}^c(X')$
we can semi-decide the condition $X'\notin P_{i_{0}}$. Therefore, whether~$p \in \mathcal L(X)$
is semi-decidable relative to an enumeration of $\mathcal{L}^{c}(X)$. In other words, from an enumeration of $\mathcal{L}^{c}(X)$ we can compute an enumeration of $\mathcal L(X)$.
\end{proof}

If we remove oracles from Theorem \ref{thm:main}, we obtain
the following characterization of the decidability of languages of
effectively closed shifts.
\begin{corollary}
Let $G$ be a finitely generated group with decidable word problem
and let $A$ be a finite alphabet. Let $X\subseteq A^{G}$ be a non-empty
effectively closed shift. The following are equivalent.
\begin{enumerate}
\item $X$ has  computable language,
\item There exists a~$\Sigma_{2}^{0}$ property~$P$ in $\mathcal{S}(A^{G})$
such that $X$ is $P$-minimal.
\end{enumerate}
\end{corollary}

\subsection{Relation with strong computable type for sets}\label{subsec:SCTforsets}

The notion of strong computable type was initially defined for sets
(see \cite{AH22c}), using the notion of copies in the \textbf{Hilbert
cube} $Q=[0,1]^{\mathbb{N}}$, and generalized the notion of
computable type for sets defined in \cite{IljazovicS18}. For a comprehensive understanding of strong computable type for sets, see Amir's PhD thesis \cite{Phdamir2024}.

A \textbf{copy} of a set $X\subseteq Q$ is the image of $X$ by a
homeomorphism $Q\rightarrow Q$ (a continuous bijection with a continuous
inverse).

$Q$ is a computable metric space, and hence there exists a computable
dense sequence $(q_{i})_{i\in\mathbb{N}}$ in $Q$. Let $(B_{i})_{i\in\mathbb{\mathbb{N}}}$ be an enumeration of
all the balls $B(q_{i},r_{j})$ with center $q_{i}$ and radius $r_{j}\in\mathbb{Q}$.

A compact set $X$ in $Q$ is \textbf{semi-computable} if the set
$\{i\in\mathbb{N}:X\cap B_{i}=\emptyset\}$ is c.e.. It is \textbf{computable}
if it is semi-computable and in addition the set $\{i\in\mathbb{N}:X\cap B_{i}\neq\emptyset\}$
is c.e.. The same notions can be defined by replacing $Q$ by any
space with a computable metric structure.

Now, let us define \textbf{strong computable type for sets}. A compact
set $X$ in $Q$ has strong computable type if for every copy $Y$
of it in $Q$, and every oracle which makes $Y$ semi-computable,
$Y$ is also computable using the oracle.

One might attempt to define strong computable type for $G$-shifts
using copies. However, this notion is very restrictive: for example, since any shift without isolated points is homeomorphic to the Cantor set, any such shift is also homeomorphic to an effectively closed shift whose language is not computable, so it does not have strong computable type.

Another possible definition involves using analogs of homeomorphisms
in symbolic dynamics, namely conjugacies. However, since conjugate
shifts preserve the decidability of languages, proving the decidability
of the language of one shift automatically implies the decidability
of the languages of all conjugate shifts. Therefore, we define strong
computable type for $G$-shifts without considering either copies
or conjugacies.

A similar result for the computability of sets, relating computable
type with minimality was proved (see \foreignlanguage{french}{\cite{AH22c,Phdamir2024})}.
We prove in Appendix \ref{Appendix:Another-proof-of} that Theorem
\ref{thm:main} can be obtained by this characterization of
strong computable type for sets, though it is not immediate.

\subsection{Relation with maximal elements in quasivarieties}\label{sec:quasivar}

In Theorem 5 of \cite{jeandel2017enumeration}, Jeandel proves a similar result for quasivarieties, which is a general framework including subshifts as well as finitely generated groups, first-order theories, etc. We compare and contrast it with our results.

Jeandel's result is a sort of dual of our Theorem~\ref{thm:main} as it applies to maximal elements and yields an enumeration reduction that is the opposite direction from strong computable type; this is only a consequence of the fact that it applies to $\mathcal{L}^c(X)$ instead of $X$. Quasivarieties correspond to (in our context) effectively closed properties that are stable under finite unions. Our result does not requires this last assumption, although note that:
\begin{itemize}
\item in the proof of the implication $1.\Rightarrow 2.$ in Theorem~\ref{thm:main}, the property $P$ is stable by union, and therefore the theorem would hold with this additional assumption;
\item properties used in our examples in Section~\ref{subsec:Application} are stable by union, so we believe these results could be proved in Jeandel's framework (indeed, examples from Sections~\ref{sec:quasiminimal} and \ref{sec:infminimal} appeared in \cite{jeandel2017enumeration}).
\end{itemize}

Additionally, Theorem 5 of \cite{jeandel2017enumeration} is not an equivalence; note that the converse given in Theorem 6 in \emph{op.cit.} is not comparable to our main theorem, and shows in the case of subshifts that the language of a subshift with strong computable type is enumeration-reducible to the language of a minimal subshift.

We are not sure of whether $\Pi_1^0$ properties and quasivarieties are equivalent in terms of expressing minimality properties (see Question~\ref{que:quasivar}). We believe that, at least, $\Pi_1^0$ properties provide an easier description of shifts with strong computable type for researchers in the symbolic dynamics community.

\subsection{Applications}\label{subsec:Application}

In this section, we provide examples of $G$-shifts that have strong
computable type based on our characterization. Most of these results are generalizations of existing results to arbitrary oracles, and a main interest of our characterization is to unify the underlying arguments.

Let $G$ be a finitely generated group with decidable word problem,
and let $A$ be a finite alphabet.

\subsubsection{Minimal shifts}\label{subsec:Minimal-subshifts}

Recall from the introduction that a non-empty shift is \textbf{minimal}
if it contains no other subshifts except itself and the empty shift.

The following result has been first stated, to the best of our knowledge,
in \cite{delvenneblondel}; it is a generalisation of a folklore result
for the finite type case.
\begin{theorem}[\cite{delvenneblondel}]
 An effectively closed minimal shift has computable language.
\end{theorem}

We prove a generalization of this theorem (relative to any oracles) based on our characterization.

\begin{proposition}
A minimal shift has strong computable type.
\end{proposition}

\begin{proof}
A minimal shift is $P$-minimal for the property $P=\mathcal S(A^G)$ (remember that the empty shift is not a member of $\mathcal S(A^G)$ by definition). This property is trivially $\Pi_1^0$, using an algorithm that never halts.
\end{proof}

\subsubsection{Entropy-minimal shifts}\label{subsec:Entropy-minimal-subshifts}

In this section we suppose in addition that the group $G$ is amenable
(we do not give the definition of being amenable as we don't explicitly
use it).
\begin{definition}
\label{def:hentropyminimal}Let $h$ be the entropy
map $\mathcal{S}(A^{G})\rightarrow\mathbb{R}^+$ that to a shift
associates its topological entropy (see Chapter 9 in \cite{kerr2017ergodic}
for more details of this notion). A $G$-shift
$X$ is called \textbf{entropy-minimal}
when every proper subshift $Y\subset X$ satisfies $h(Y)<h(X)$.
\end{definition}

The following result is stated in a recent unpublished work by Carrasco-Vargas,
Herrera Nunez and Sablik that appeared in Chapter 6 in Carrasco-Vargas's PhD
Thesis \cite{Phdnicanor2024}.
\begin{theorem}[\cite{Phdnicanor2024}]
An entropy-minimal SFT whose entropy is a computable real number
has computable language.
\end{theorem}

A real number is \textbf{left-computable} if there exists a computable
increasing sequence of rational numbers converging to it.

In Proposition \ref{prop:An-entropy-minimal-subshift}, we generalize
this result in several ways: from SFTs to effectively closed shifts, from computable entropies to left-computable entropies, and finally from having computable languages to having strong computable type, based on our characterization in Theorem \ref{thm:main}.
\begin{proposition}
\label{prop:An-entropy-minimal-subshift}An entropy-minimal shift $X$
whose entropy is a real number that is left-computable relative to $\mathcal L^c(X)$ has strong computable
type.
\end{proposition}

\begin{proof}
Let $X$ be an entropy-minimal shift with $q = h(X)$. Take the property $P=h^{-1}([q,+\infty))$. Relative to $\mathcal L^c(X)$, $P$ is $\Pi_{1}^{0}$ because
$q$ is left-computable and $h$ is upper semi-computable (combine
Proposition 6.17 in \cite{Phdnicanor2024} with Fact \ref{fact:Sp01}).
Clearly, $X$ is $P$-minimal, so by Theorem \ref{thm:main} it has strong computable type.
\end{proof}

\begin{remark}
The previous proof also works when $h(X)$ is not left-computable, under the assumption that $\sup_{Y\subsetneq X} h(Y) < h(X)$. In this case, pick $q$ to be any left-computable number in the interval $[\sup_{Y\subsetneq X} h(Y), h(X)]$.
\end{remark}

\paragraph*{Entropy-minimality alone is not sufficient}

We give an example of a shift in $\{0,1\}^{\mathbb{Z}}$ that is entropy-minimal
and does not have strong computable type.
\begin{definition}
\label{def:sturmian}Let $\alpha\leq\beta$ be two real numbers. Define
the shift $X_{[\alpha,\beta]}\subset\{0,1\}^{\mathbb{Z}}$ by forbidding
the set of finite patterns 
\[
\{w\in\{0,1\}^{n}:\#_{1}w<\lfloor\alpha n\rfloor\text{ or }\#_{1}w>\lceil\beta n\rceil\},
\]
where $\#_{1}w$ is the number of occurrences of the symbol $1$ in
$w$. $X_{\alpha}=X_{[\alpha,\alpha]}$ is known as the Sturmian shift
of slope $\alpha$.
 \end{definition}

See \cite[Chapter 2]{lothaire2002algebraic} for more properties of
Sturmian configurations and Sturmian shifts. In particular, the language
of a Sturmian shift with irrational slope contains exactly $n+1$
different patterns of length $n$.
\begin{proposition}
\label{Prop:entropyminimal} The shift $X_{[0,\alpha]}$ is entropy-minimal and, for some right-computable value of $\alpha$, does not have strong computable type.
\end{proposition}

\begin{proof}
The shift $X_{[0,\alpha]}$ is \emph{strongly irreducible}, that is, for two patterns $u,v\in \mathcal L(X_{[0,\alpha]})$, we have $u0^kv\in \mathcal L(X_{[0,\alpha]})$ for some constant $k = \lceil \frac2\alpha\rceil$. Indeed, check that $\#_{1}u0v= \#_{1}u+\#_{1}v \leq \lceil\alpha |u| \rceil + \lceil\alpha |v|\rceil \leq \alpha(|uv|+2) \leq \lceil\alpha (|uv|+k) \rceil$, and the same argument applies to any subpattern of $u0^kv$. A strongly irreducible shift is entropy-minimal: see Corollary~4.7 in \cite{fiorenzi2004semi}.

If $\alpha$ is a right-computable real number (namely, $-\alpha$
is left-computable), then $X_{[0,\alpha]}$ is an effectively closed shift
since the set $\{w\in\{0,1\}^{n}:\#_{1}w>\lceil\alpha n\rceil\}$
is c.e.. In particular, $\mathcal L(X_{[0,\alpha]})$ is co-computably enumerable.

By an argument that can be found e.g. in the proof of Theorem~3.6 in \cite{gangloff2016effect}, for a
strongly irreducible shift $Y$, from the number of patterns of size $n$ in $\mathcal L(Y)$ we can compute
an approximation of $h(Y)$ with a known error rate.
In our case, having a right-approximation of number of patterns of size $n$ in $\mathcal L(X_{[0,\alpha]})$, we obtain that
$h(X_{[0,\alpha]})$ is right-computable.

Now assume that $X_{[0,\alpha]}$ has computable language.
Compute all patterns of length $n$ in $\mathcal{L}(X_{[0,\alpha]})$
and count the maximum number of symbols $1$ that appear in a pattern:
this number is $\lfloor\alpha n\rfloor$, from which we get an approximation
of $\alpha$ up to error $1/n$. Therefore, $\alpha$ is a computable
real number in this case.

It follows that $X_{[0,\alpha]}$, for $\alpha$ right-computable
but not computable, is entropy-minimal, effectively closed and its language
is not computable, so it does not have strong computable type.
\end{proof}

\subsubsection{P-isolated shifts}

As in the previous section, studying isolated points is motivated by the unpublished work
of Carrasco-Vargas, Herrera Nunez and Sablik that appeared in \cite{Phdnicanor2024}.

In their work they find an alternative proof that a minimal SFT has computable language,
using the fact that it is isolated in $\mathcal{S}(A^{G})$.

Similarly, they prove that an entropy-minimal SFT with computable
real entropy $q$ has computable language, using the fact that it is isolated in $h^{-1}([q,+\infty))$, see Definition \ref{def:hentropyminimal}.

The common property between $\mathcal{S}(A^{G})$ and $h^{-1}([q,+\infty))$
is that they are $\Pi_{1}^{0}$ (see Fact \ref{fact:Sp01}
in the Appendix for $\mathcal{S}(A^{G})$). In general, isolated points in $\Pi_1^0$ classes are computable: see Fact~2.14 in \cite{DOWNEY_MELNIKOV_2023}. It motivates us to generalize the result as follows.

\begin{proposition}
\label{prop:isolated} Let $X\in\mathcal{S}(A^{G})$ be a non-empty
shift and $P$ be a property in $\mathcal{S}(A^{G})$
which is $\Sigma_{2}^{0}$ relative to $\mathcal L^c(X)$.  If $X$ is isolated in $P$, then $X$ has strong computable
type.
\end{proposition}

\begin{proof}
Since $X$ is isolated in $P$, we can find a cylinder $[p]$ such that
$X \cap [p] \neq \emptyset$ and $Y\in P \Rightarrow Y\cap [p] = \emptyset$. The property $C = \{Y : Y\cap [p] \neq \emptyset\}$ is both $\Sigma_{1}^{0}$ and $\Pi_{1}^{0}$.
Let $P'=P\cap C=\{X\}$. $P'$ is $\Sigma_{2}^{0}$ relative to $\mathcal L^c(X)$ because it is the intersection of a property that is $\Sigma_{2}^{0}$ relative to $\mathcal L^c(X)$ with a $\Sigma_{1}^{0}$
property. Clearly, $X$ is $P'$-minimal, hence it has strong computable type.
\end{proof}

\subsubsection{Periodic-minimal shifts}
\begin{definition}
The \emph{orbit} $\Orb(x)$ of a configuration $x\in\A^{G}$ is the
closure of the set $\{g\cdot x:g\in G\}$. A (strongly) periodic configuration
is a configuration $x$ such that the number of elements $\#\Orb(x)$
is finite.
\end{definition}

To a shift $X$ we associate the vector $\Per(X)=(\Per_{i}(X))_{i\in_{\N}}$,
where $\Per_{i}(X)$ is the number of periodic points in $X$ with orbit
size $i$ or less. This is a classical conjugacy invariant for shifts,
usually presented under the form of a Zeta function; see e.g. \cite{lind1996zeta}
for more information on this object.

A shift $X$ is \textbf{period-minimal} if, for any subshift $Y\subsetneq X$,
$\Per_{i}(Y)<\Per_{i}(X)$ for some $i\in\N$. This corresponds to
minimality for the $\Pi_{1}^{0}$ property $P_{X}=\{Y:\forall i,\Per_{i}(Y)\geq\Per_{i}(X)\}$;
therefore a period-minimal shift has strong computable type.
\begin{proposition}
A shift is period-minimal if, and only if, it has dense periodic points.
These shifts have strong computable type. 
\end{proposition}

\begin{proof}
If $X$ is period-minimal, then the subshift obtained from $X$ by forbidding any pattern $p\in \L(X)$ has strictly less periodic points.
This means that any pattern $p \in \mathcal L(X)$ appears in some periodic point of $X$, that is, $[p]\cap X$ contains a periodic point.
Since cylinders are a basis of the topology, periodic points are dense in $X$. Conversely, if periodic points are dense in $X$,
then any strict subshift $Y\subsetneq X$ has strictly less periodic points.
\end{proof}
We obtained hence a stronger version of a classical result:
\begin{theorem}[\cite{kitchens2006periodic}]
\label{thm:periodic} An effectively closed shift with dense periodic points
has computable language.
\end{theorem}

\subsubsection{Quasi-minimal shifts}\label{sec:quasiminimal}

Quasi-minimal shifts as introduced by Salo \cite{salo2017decidability} are shifts with finitely many distinct subshifts.Let
\begin{theorem}[\cite{salo2017decidability}, Theorem 9.]
\label{thm:quasiminimal} An effectively closed quasi-minimal shift has computable
language.
\end{theorem}

We strenghten this result by proving the following.
\begin{proposition}
Quasi-minimal shifts have strong computable type.
\end{proposition}

\begin{proof}
For a given quasi-minimal shift $X$, denote $X_{1}\dots X_{n}$ its
distinct subshifts and fix $p_{i}\in\L(X)\backslash\L(X_{i})$ for
all $1\leq i\leq n$. $X$ is then minimal for the $\Pi_{1}^{0}$
property $P_{X}=\{Y:\forall i\leq n,p_{i}\in\L(Y)\}$.
\end{proof}
In \emph{op.cit}, Salo studies a different notion, initially introduced
in \cite{delvenneblondel}, which is having finitely many distinct
\emph{minimal} subshifts. We do not believe that such shifts must
have strong computable type.

\subsubsection{Infinite-minimal shifts}\label{sec:infminimal}

It is well-known \cite[Theorem 3.8]{ballier2008structural} that a
shift is finite if and only if it contains only strongly periodic
points. We consider the property for a shift $X$ to be \textbf{infinite-minimal},
that is, $|X|=\infty$ and all subshifts $Y\subsetneq X$ are finite. 
Equivalently, such a shift is minimal for the property of containing a non-periodic configuration. This property was called \emph{just-infinite} in \cite{jeandel2017enumeration}.
The property of being infinite is known to be $\Pi_{0}^{1}$ in $\Z^{2}$
but not in $\Z^{d}$, for $d>2$ (see \cite{callard2022aperiodic}),
so we do a direct proof.
\begin{proposition}
An infinite-minimal shift has strong computable type.
\end{proposition}

\begin{proof}
If there are finitely many such $Y$'s, then $X$ is quasi-minimal
and we conclude by Theorem~\ref{thm:quasiminimal}. Otherwise, $X$
contains infinitely many periodic points. If there is a pattern $p$
that appears in no periodic point, the subshift obtained from $X$
by forbidding $p$ contains all periodic points of $X$, so it is
infinite, which contradicts infinite-minimality. We conclude that periodic points are dense in $X$, which
is enough by Theorem~\ref{thm:periodic}.
\end{proof}

\subsubsection{Full extensions}

Jeandel and Vanier proved the following characterization.
\begin{theorem}[\cite{jeandel2019characterization}]
\foreignlanguage{american}{ A shift $X$ in $A^{\mathbb{Z}^{d}}$
has computable language if and only if it is the \emph{projective subdynamics} of a minimal effectively closed
shift $Y$ (possibly with larger alphabet and dimension); in other words,
$\mathcal{L}(X)=\mathcal{L}(Y)\cap \mathcal L(A^{\mathbb{Z}^{d}})$, which are patterns $p : F\to A \in\mathcal L(Y)$ such that $F\subset \mathbb{Z}^{d}$.}
\end{theorem}

We show that if $X$ is the projective subdynamics of a minimal shift $Y$ then it has strong computable type, which corresponds to one direction of the previous result relative to an oracle. It is possible that the proof of \cite{jeandel2019characterization} can also be generalised for the other direction.

Assume that $Y\subset B^{\mathbb Z^{d'}}$ with $A\subset B$ and $d\leq d'$. We remark that in the previous theorem, we have a $\Pi_{1}^{0}$ property $P$ for which $X$ is minimal:
\begin{align*}
P & =\{Z\in\mathcal{S}(A^{\mathbb{Z}^{d}}):Y\backslash\bigcup_{g\in\mathbb{Z}^{d'}}\bigcup_{w\in\mathcal{L}^{c}(Z)}g[w]\neq\emptyset\}.
\end{align*}
Notice that since $Z$ is a subshift of $A^{\mathbb{Z}^{d}}$, $\mathcal{L}^{c}(Z)$ contains patterns $F\to A$ with $F\subset \mathbb Z^d$, which are seen as patterns $F\to B$ with $F\subset \mathbb Z^{d'}$.

Let us prove that $X$ is $P$-minimal. $Y\backslash\bigcup_{g\in\mathbb{Z}^{d'}}\bigcup_{w\in\mathcal{L}^{c}(X)}g[w]=Y\neq\emptyset$
so $X\in P$. Now let $X'$ be a proper subshift in $X$: this means
that there exists some pattern $p\in \mathcal L(X)\setminus \mathcal L(X')$. By assumption, $p\in\mathcal L(Y)$. Since $Y$ is minimal, $Y\backslash\bigcup_{g\in\mathbb{Z}^{d'}}\bigcup_{w\in\mathcal{L}^{c}(X')}g[w]=\emptyset$
hence $X'\notin P$.

As $Y$ is effectively closed and minimal, it has computable language
(see Section \ref{subsec:Minimal-subshifts}), which makes the condition
$Y\backslash\bigcup_{g\in\mathbb{Z}^{d'}}\bigcup_{w\in\mathcal{L}^{c}(Z)}g[w]=\emptyset$ semi-decidable
given an enumeration of $\mathcal{L}^{c}(Z)$. Hence $P$ is a $\Pi_{1}^{0}$ property.

\subsection{Invariance under transformations}

Strong computable type is preserved under products and factors. However,
it is not preserved under unions and intersections.

\subsubsection{Factors and computable morphisms}

Let $A$ and $B$ be two finite alphabets. A factor $F:\mathcal{S}(A^{G})\rightarrow\mathcal{S}(B^{G})$
is a surjective morphism between $G$-shifts that commutes with the
shift action. It sends shifts with computable languages in $\mathcal{S}(A^{G})$
to shifts with computable languages in $\mathcal{S}(B^{G})$ because
it is a computable surjective function. Using oracles, it sends shifts
with strong computable type to shifts with strong computable type.
More generally, an image of a shift with strong computable type by
a computable morphism has strong computable type.

\subsubsection{Products}

In contrast to the case of sets, where products do not always preserve strong computable type (see \cite{AMIR2024109020}),
strong computable type for shifts is preserved under finite products. 

Let $A$ and $B$ be two finite alphabets. Let $X$ be a shift in
$\mathcal{S}(A^{G})$ and $Y$ be a shift in $\mathcal{S}(B^{G})$,
the product $X\times Y$ is a shift in $\mathcal{S}(A^{G})\times\mathcal{S}(B^{G})\subseteq\mathcal{S}(A^{G}\times B^{G})\equiv\mathcal{S}((A\times B)^{G})$.
\begin{proposition}
If two shifts $X$ and $Y$ have strong computable type, then $X\times Y$
has strong computable type.
\end{proposition}

\begin{proof}
Assume that $X \subset A^G$ and $Y\subset B^G$. A pattern $w : F\to A\times B \in \mathcal{L}^{c}(X\times Y)$ can be seen as the product of two patterns $u: F\to A$ and $v: F\to B$ such that either $u\in\mathcal{L}^{c}(X)$ or $v\in\mathcal{L}^{c}(Y)$.
For a pattern $p : F\to A$, $p\in\mathcal{L}^{c}(X)$ is equivalent to the fact that all the pairs $(p,q)$ with $q : F\to B$ are in $\mathcal{L}^{c}(X\times Y)$. Indeed, if $p\in\mathcal{L}(X)$, then only pairs such that $q\in \mathcal{L}^{c}(Y)$ will be enumerated and $\mathcal{L}(Y)\neq\emptyset$.
Therefore, given an enumeration of $\mathcal{L}^{c}(X\times Y)$
one can enumerate the elements of $\mathcal{L}^{c}(X)$.

By symmetry, one can enumerate the elements of $\mathcal{L}^{c}(Y)$.
As $X$ and $Y$ have strong computable type, $\mathcal{L}(X)$ and
$\mathcal{L}(Y)$ are computable. Hence $\mathcal{L}(X\times Y)=\mathcal{L}(X)\times\mathcal{L}(Y)$
is computable and $X\times Y$ has strong computable type.
\end{proof}

\subsubsection{Unions and intersections}
\begin{proposition}
There exists two shifts $X,Y$ that have strong computable type such
that $X\cup Y$ does not have strong computable type. Similarly, there exists two shifts $X'$ and $Y'$ that have strong computable type such
that $X'\cap Y'$ does not have strong computable type.
\end{proposition}

\begin{proof}
Consider $X_{[\alpha,\alpha+1/2]}$ and $X_{[\beta-1/2,\beta]}$ for
some real numbers $\alpha<\beta-1/2<\alpha+1/2<\beta$ (see Definition
\ref{def:sturmian}). It is clear that $X_{[\alpha,\alpha+1/2]}\cup X_{[\beta-1/2,\beta]}=X_{[\alpha,\beta]}$.
If $\alpha$ is left-computable and $\beta$ is right-computable (namely,
$-\beta$ is left-computable), but one of them is not computable,
then $X_{[\alpha,\beta]}$ is effectively closed but its language is not computable
by the same argument as in Proposition~\ref{Prop:entropyminimal},
so it does not have strong computable type. However, from an enumeration
of $\L^{c}(X_{[\alpha,\alpha+1/2]})$ and by counting the number of
symbols $1$ in each, it is not hard to compute an upper approximation
of $\alpha$ and a lower approximation of $\alpha-1/2$, and therefore
to compute arbitrarily good approximations of $\alpha$ with known
error, from which it is straightforward to compute the language of
$X_{[\alpha,\alpha+1/2]}$. It follows that $X_{[\alpha,\alpha+1/2]}$
has strong computable type. The case of $X_{[\beta-1/2,\beta]}$ is
symmetric. 

For the intersection, apply a similar argument to the same shifts when $\beta-1/2<\alpha<\beta<\alpha+1/2$.
\end{proof}

The next proposition shows that strong computable type is conserved under disjoint unions. It is not enough that the intersection $X \cap Y$ is simple (in a computational sense) since, in the previous proof, we had $X_{[\alpha,\alpha+1/2]}\cap X_{[\beta-1/2,\beta]} = X_{[\beta-1/2,\alpha+1/2]}$ which has computable language.

\begin{proposition}
Let $X$ and $Y$ be two shifts with strong computable type such that $X\cap Y = \emptyset$. Then $X\cup Y$ have strong computable type.
\end{proposition}

\begin{proof}
In this proof we assume that the underlying group $G$ is generated by the set of generators $S$ and we denote $B_S(n)=\{w_G : w\in S^{\leq n}\}$.

Since $X\cap Y = \emptyset$, a compactness argument shows that there is a $N\in \N$ such that, if $p \in A^{B_S(n)}$ for $n>N$,  $[p]\cap X = \emptyset$ or $[p]\cap Y = \emptyset$. Indeed, if that was not the case, we would get a sequence of patterns with increasing support from which we could extract by compacity an element of $X\cap Y$.

Denote by $E_Y$ the set of patterns defined as follows. For any pattern $p$ of support $F$,
\begin{itemize}
\item if $F \subset B_S(N)$, $p\in E_Y$ if and only if $[p]\cap X = \emptyset$;
\item otherwise, let $n$ be the smallest number such that $F \subset B_S(n)$. Then $p\in E_Y$ if and only if, for any pattern $p'$ of support $B_S(N)$ that extends $p$ (that is, $p'|_{F\cap B_S(N)}=p|_{B_S(N)}$), $[p'|_{B_S(N)}]\cap Y\neq \emptyset$.
\end{itemize}

Checking whether $p\in E_Y$ requires finitely many checks of the form $[q]\cap X = \emptyset$ or $[q]\cap Y \neq \emptyset$ for a pattern $q$ of support included in $B_S(N)$. There are finitely many such patterns, so $E_Y$ is a computable set.

Now assume that we receive as input an enumeration of $\L^c(X\cup Y) = \L^c(X)\cap \L^c(Y)$. We show that $\L^c(X\cup Y) \cup E_Y = \L^c(X)$. It is clear that $E_Y\subset\L^c(X)$ by definition of $N$ and $E_Y$. Furthermore, $\L(Y)\cap \L^c(X) \subset E_Y$; in fact the first point  includes all "small" patterns in $\L^c(X)$ and the second point includes all "large" patterns in $\L(Y)$.

 We compute an enumeration of $\L^c(X\cup Y) \cup E_Y = \L^c(X)$. Since $X$ has strong computable type, we can compute a enumeration of $\L(X)$. Doing similarly for $Y$, we compute an enumeration of $\L(Y)$. From this we obtain an enumeration of $\L(X\cup Y) = \L(X) \cup \L(Y)$, so $X\cup Y$ has strong computable type.
\end{proof}

The previous proof relies on the fact that, in $A^G$, two disjoint shifts  (or more generally two disjoint closed sets) $X$ and $Y$ are computably separable; $E_Y$ plays the role of a computable superset of $Y$ that is disjoint from $X$.

\section{Discussion, generalizations and future directions}\label{sec:Generalization,-conclusion-and}

Several results in this article can be generalized to recursively
presented groups.

The notion of strong computable type for shifts can be extended to
pairs of shifts, consisting of a shift and a subshift of it, in the
same way as the notion of strong computable type for pairs of sets.
Similar results can then be proved, and we will address this in a
subsequent article that extends this one.

Our results for strong computable type for shifts can be analogously
extended to strong computable type for sets. For example, $P$-isolated
sets, where $P$ is a topological $\Sigma_{2}^{0}$ invariant, have
strong computable type.
\begin{question}
Let us consider further results on the computable type of sets. Can
we obtain analogous results for shifts, and vice versa?
\end{question}

The results here also motivate an independent research direction:
the study and characterization of the descriptive complexity of properties
of shifts, similar to how such properties have been studied for sets
in \cite{AH22c,AMIR2025103611}.
\begin{question}
Can we characterize $\Sigma_{2}^{0}$ and $\Pi_{1}^{0}$ properties
of shifts?
\end{question}

Motivated by the more general approach of \cite{jeandel2017enumeration}, we would like to better understand the relationship between Jeandel's and our results and find a unifying argument that can be developed using computable analysis, as we have done here for shifts.

\begin{question}\label{que:quasivar}
Is it possible to obtain results based on minimality similar to Theorem~\ref{thm:main} for other theories, such as group theory and combinatorics, in a unifying framework based on computable analysis?
\end{question}

\section*{Acknowledgements}  

	The authors are grateful to anonymous reviewers, as well as to Pierre Guillon and Guillaume Theyssier for discussions and numerous remarks that helped to significantly improve this article. We particularly thank Pierre Guillon for pointing out links with results from \cite{jeandel2017enumeration}. \\
	We gratefully acknowledge funding from the ANR Project IZES-ANR-22-CE40-0011 (Inside Zero Entropy Systems) for this work.

\bibliographystyle{plain}
\bibliography{biblio}

\appendix

\section[Topology on A\^{}G and its hyperspace]{Topology on $A^{G}$ and its hyperspace}\label{Appendix:Topology-on}

\subsection{Topology on $A^{G}$}\label{subsec:Topology-on-1}

To define $G$-shifts we need to induce $A^{G}$ with a topology as
follows (see \cite{ceccherini2010cellular}).
\begin{definition}
Let $G$ be a group and $A$ be an alphabet. We endow $A^{G}$ with
the \textbf{pro-discrete} topology obtained by inducing $A$ with
the discrete topology (each point in $A$ is open and closed (clopen))
and then taking the product topology. 

If $G=\{g_{i}:i\in\mathbb{N}\}$ is countable, this topology is metrizable
using the metric 
\[
d(x,y)=\inf\{\frac{1}{n}:n\in\mathbb{N},\text{ and }x(i)=y(i)\text{ for all }i\leq n\}.
\]

If $G$ is finitely generated and $S$ is a generating set, they induce
a \textbf{word length} $|\cdot|_{S}:G\rightarrow\mathbb{N}$ such
that if $g\in G$, $|g|_{S}$ is the length of the shortest word in
$S^{*}$ with $g=w_{G}$. Consequently, a \textbf{word metric} $d_{S}$
on $G$ can be defined as follows: for every $(g,h)\in G^{2}$, $d_{S}(g,h)=|g^{-1}h|_{S}$.
In this case another metric for the pro-discrete topology on $A^{G}$
is 
\[
d(x,y)=\inf\{\frac{1}{n}:n\in\mathbb{N},\text{ and }x(g)=y(g)\text{ for all }|g|_S\leq n\}.
\]
\end{definition}

\begin{fact}[Section 3.3. in \cite{barbieri2024effective}]
\label{fact:AGcomputablemetric}
Let $A$ be an alphabet and $G$ a finitely generated group with decidable word problem. The space $A^{G}$ can be endowed with a computable
metric structure corresponding to the distance $d$ defined above: that is, there exists a dense
sequence $(x_{i})_{i\in\mathbb{N}}\subseteq A^{G}$ which is uniformly
computable (i.e. there exists an algorithm which given any pair $(i,j)\in\mathbb{N}^{2}$
and any $n\in\mathbb{N}$ computes some $\alpha\in\mathbb{Q}$ such
that $|d(x_{i},x_{j})-\alpha|<1/n$).

\end{fact}

\subsection[The hyperspace of A\^{}G]{The hyperspace of $A^{G}$}\label{subsec:The-hyperspace-of}

Let $G$ be a finitely generated group with decidable word problem
and let $A$ be a finite alphabet. We denote by $\mathcal{K}(A^{G})$
the \textbf{hyperspace} of $A^{G}$ i.e. the space of all non-empty
compact subsets of $A^{G}$. It can be endowed with the induced Hausdorff
metric (see Section 2.3. in \cite{Phdamir2024}) which induces also
a computable metric structure (see \cite{Iljazović2021}) from the
computable metric structure defined in Fact \ref{fact:AGcomputablemetric}.
This metric structure induces a topology which is called Vietoris
topology, see Section 2.3. in \cite{Phdamir2024}.

A \textbf{property }is a subset of $\mathcal{K}(A^{G})$.

\subsubsection{Descriptive complexity for properties in $\mathcal{K}(A^{G})$ and
$\mathcal{S}(A^{G})$}\label{subsec:Descriptive-complexity-for}

Descriptive complexity for properties in $\mathcal{K}(A^{G})$ can
be defined in the same way as for properties in $\mathcal{S}(A^{G})$
in Definition \ref{def:The-descriptive-complexity}. As $\mathcal{S}(A^{G})$
is a property in $\mathcal{K}(A^{G})$ it is not hard to see its descriptive
complexity.
\begin{fact}
\label{fact:Sp01}$\mathcal{S}(A^{G})$ is a $\Pi_{1}^{0}$
property in $\mathcal{K}(A^{G})$. 
\end{fact}

\begin{proof}
Let $X\in\mathcal{K}(A^{G})$ be a set, given an enumeration of $\mathcal{L}(X)$
and its complement, we can semi-decide whether $X\notin\mathcal{S}(A^{G})$
i.e. the fact that $X$ is not shift-invariant. Indeed, eventually
the machine will enumerate some pattern $p$ and some $g\in G$ such
that $[p]\cap X\neq\emptyset$ and $g[p]\cap X=\emptyset$.
\end{proof}
In Section \ref{subsec:Application}, the properties we define in
$\mathcal{S}(A^{G})$ can be defined more generally in $\mathcal{K}(A^{G})$,
for instance the property of being non-empty is $\Pi_{1}^{0}$ in
$\mathcal{K}(A^{G})$.

Now, let us discuss the relation between topologies on $\mathcal{K}(A^{G})$
and $\mathcal{S}(A^{G})$ and the descriptive complexity of properties.

As $\mathcal{K}(A^{G})$ is a computable metric space, let $(k_{i})_{i\in\mathbb{N}}$
be a computable dense sequence in $\mathcal{K}(A^{G})$. One can enumerate
all the balls of the form $B(k_{i},r_{j})$ with center $k_{i}$ and
radius $r_{j}\in\mathbb{Q}$, let $(B_{k})_{k\in\mathbb{\mathbb{N}}}$
be such an enumeration. $\mathcal{S}(A^{G})$ can then be endowed with
the induced topology as a subspace of $\mathcal{K}(A^{G})$.

A property is \textbf{effectively open} if there is some c.e. $I\subseteq\mathbb{N}$,
such that $P=\bigcup_{i\in I}B_{i}$.

A property is \textbf{effectively closed} if it is the complement
of an effectively open property.

Here is a reformulation of descriptive complexity using effective
topology.
\begin{fact}
\label{fact:effclo<>p01}A property in
$\mathcal{K}(A^{G})$ is $\Sigma_{1}^{0}$ iff it is effectively open
in $\mathcal{K}(A^{G})$. It is $\Pi_{1}^{0}$ if it is effectively
closed in $\mathcal{K}(A^{G})$. The same is true for properties in $\mathcal{S}(A^{G})$.
\end{fact}

\subsubsection{Descriptive complexity and upper Vietoris topology}\label{subsec:Another-definition-of}

We explained in Fact \ref{fact:effclo<>p01}
that being $\Pi_{1}^{0}$ in $\mathcal{K}(A^{G})$ is equivalent to
being effectively closed in the Vietoris topology induced by the Hausdorff
metric. The same is true for properties
in $\mathcal{S}(A^{G})$.

There is a stronger notion that corresponds to $P$ being effectively closed in the upper Vietoris topology.
This is equivalent to the existence of an algorithm that, for
every $X\in\mathcal{K}(A^{G})$, given an enumeration
of $\mathcal{L}^{c}(X)$, halts if and only if $X\notin P$. The difference with Vietoris topology is that the algorithm is given as input only the enumeration of the complement.
The same is true for properties in $\mathcal{S}(A^{G})$.

The reason why we could assume this stronger notion in the proof of Theorem \ref{thm:main}
is the following fact, see Proposition 4.4.4. in Amir's PhD thesis
\cite{Phdamir2024}. 
\begin{fact}
\label{fact:PinV<>PinupV}If
$P$ is a $\Sigma_{2}^{0}$ property in the Vietoris topology, then
there exists some property $P'$ which is $\Sigma_{2}^{0}$ in the
upper Vietoris topology and such that they induce the same minimal
elements, i.e. $X$ is $P$-minimal if and only if $X$ is $P'$-minimal.
\end{fact}

\section{Another proof of Theorem \ref{thm:main}}\label{Appendix:Another-proof-of}

In \cite{AH22c}, a characterization for sets with strong computable
type was provided. Our Theorem \ref{thm:main} is motivated
by this result and, in fact, can be derived from it (though not directly).
Let us explain why.

Let $G$ be a finitely generated group with  decidable word problem,
and let $A$ be a finite alphabet.

First, let us state the result for sets, reformulated without using
copies (see Theorem 4.7. in \cite{AH22c}).
\begin{theorem}[A characterization of strong computable type for set]
\label{thm:SCTforsets}Let $X$ be a compact
subset in $A^{G}$. The following are equivalent.
\begin{enumerate}
\item For every oracle $O\subset\mathbb{N}$, if $X$ is semi-computable
relative to $O$, then it is computable relative to $O$.
\item There exists a~$\Sigma_{2}^{0}$ property~$P$ in $\mathcal{K}(A^{G})$
such that $X$ is $P$-minimal.
\end{enumerate}
\end{theorem}

\begin{fact}
\label{fact:Effective-semicomputable}Effectively closed $G$-shifts are exactly
the semi-computable compact shift-invariant subsets of $A^{G}$. (This is not straightforward; it is a theorem, see \cite{AUBRUN201735}, since
for effectively closed shifts we enumerate words, and for semi-computable subsets,
we enumerate balls.)
\end{fact}

The results from a recent, yet unpublished work by Carrasco-Vargas,
Herrera Nunez and Sablik, (see Proposition 6.12. in \cite{Phdnicanor2024})
imply that $G$-shifts with computable languages are exactly the computable
shift-invariant compact subsets of $A^{G}$. To show that, one needs to see
(semi-)computable compact subsets of $A^{G}$ as computable points in the
(upper) Vietoris topology, as explained in Section 2.3. in \cite{Phdamir2024}.

We claim that this result can be relativized in the following way.
\begin{proposition}
\label{prop:decidable-computable}Let $O\subset\mathbb{N}$ be an
oracle, a $G$-shift has computable language relative to $O$ iff
it is a computable point relative to $O$ in the Vietoris topology.
\end{proposition}

Let us see how to obtain to Theorem \ref{thm:main} from Theorem \ref{thm:SCTforsets}.

Suppose we have $1.$ in Theorem \ref{thm:main} for some shift $X$. By Proposition \ref{prop:decidable-computable} and the relative version of Fact \ref{fact:Effective-semicomputable}, we have $1.$ in Theorem \ref{thm:SCTforsets}, so we find a~$\Sigma_{2}^{0}$ property~$P$ in $\mathcal{K}(A^{G})$
such that $X$ is $P$-minimal. Hence, $X$ is minimal for the property $P'=P \cap \mathcal{S}(A^{G})$ which is $\Sigma_{2}^{0}$ in $\mathcal{S}(A^{G})$ ($2.$ in Theorem \ref{thm:main}).
For the other direction, remark that a $\Sigma_{2}^{0}$ property in $\mathcal{S}(A^{G})$ is also $\Sigma_{2}^{0}$ in $\mathcal{K}(A^{G})$, as $\mathcal{S}(A^{G})$ is $\Pi_1^0$ in $\mathcal{K}(A^{G})$ by Fact \ref{fact:Sp01}.
\end{document}